\documentclass[
    reprint,
    superscriptaddress,
    amsmath,amssymb,
    aps,
    prx,
    longbibliography
]{revtex4-1}

\usepackage{braket}
\usepackage{graphicx}
\usepackage{amsmath}
\usepackage{amssymb}
\usepackage{float}
\usepackage[dvipsnames]{xcolor}
\usepackage[colorlinks=true,urlcolor=blue,linkcolor=blue,citecolor=blue]{hyperref}
\usepackage{cleveref}
\usepackage[margin=0.75in]{geometry}
\usepackage{physics}

\crefname{equation}{Eq.}{Eqs.}
\Crefname{equation}{Equation}{Equations}
\crefname{figure}{Fig.}{Figs.}
\Crefname{figure}{Figure}{Figures}
\crefname{section}{Sec.}{Sects.}
\Crefname{section}{Section}{Sections}
\crefname{table}{Table}{Tables}
\crefname{appendix}{Appendix}{Apps.}
\Crefname{appendix}{Appendix}{Apps.}

\newcommand{\wdrive}{\omega_d}

\newcommand{\ncrit}{n_\textrm{crit}}

\begin{document}

\author{Ross Shillito}
\affiliation{Institut Quantique and D\'epartement de Physique, Universit\'e de Sherbrooke, Sherbrooke J1K 2R1 QC, Canada}
\affiliation{Sandbox@Alphabet, Mountain View, CA 94043, USA}

\author{Alexandru Petrescu}
\affiliation{Institut Quantique and D\'epartement de Physique, Universit\'e de Sherbrooke, Sherbrooke J1K 2R1 QC, Canada}
\affiliation{Laboratoire  de Physique de l’Ecole Normale Supérieure, Mines-Paristech, CNRS, ENS-PSL, Inria,  Sorbonne Université, PSL Research University, Paris, France}

\author{Joachim Cohen}
\affiliation{Institut Quantique and D\'epartement de Physique, Universit\'e de Sherbrooke, Sherbrooke J1K 2R1 QC, Canada}

\author{Jackson Beall}
\affiliation{Sandbox@Alphabet, Mountain View, CA 94043, USA}

\author{Markus Hauru}
\affiliation{Sandbox@Alphabet, Mountain View, CA 94043, USA}

\author{Martin Ganahl}
\affiliation{Sandbox@Alphabet, Mountain View, CA 94043, USA}

\author{Adam G.M. Lewis}
\affiliation{Sandbox@Alphabet, Mountain View, CA 94043, USA}

\author{Guifre Vidal}
\affiliation{Sandbox@Alphabet, Mountain View, CA 94043, USA}
\affiliation{Canadian Institute for Advanced Research, Toronto M5G 1M1 ON, Canada}

\author{Alexandre Blais}
\affiliation{Institut Quantique and D\'epartement de Physique, Universit\'e de Sherbrooke, Sherbrooke J1K 2R1 QC, Canada}
\affiliation{Canadian Institute for Advanced Research, Toronto M5G 1M1 ON, Canada}

\title{Dynamics of Transmon Ionization}

\begin{abstract}
Qubit measurement and control in circuit QED rely on microwave drives, with higher drive amplitudes ideally leading to faster processes. However, degradation in qubit coherence time and readout fidelity has been observed even under moderate drive amplitudes corresponding to few photons populating the measurement resonator. Here, we numerically explore the dynamics of a driven transmon-resonator system under strong and nearly resonant measurement drives, and find clear signatures of transmon ionization where the qubit escapes out of its cosine potential. Using a semiclassical model, we interpret this ionization as resulting from resonances occurring at specific resonator photon populations. We find that the photon populations at which these spurious transitions occur are strongly parameter dependent and that they can occur at low resonator photon population, something which may explain the experimentally observed degradation in measurement fidelity.
\end{abstract}

\date{\today}

\maketitle

\section{Introduction}

Dispersive readout in circuit QED is realized by driving a measurement resonator coupled to the qubit \cite{wallraff2005a}. In principle, increasing the drive amplitude, and thereby the resonator photon population, increases the measurement rate something which is expected to lead to fast, high-fidelity and quantum non-demolition (QND) readout \cite{Blais2021}. However, experimentally the fidelity and QND character of the measurement of transmon qubits \cite{Koch_2007_Transmon} is often observed to decrease beyond a photon number threshold which can be as small as a few photons~\cite{Walter2017,Minev2019}. 
Perturbative models have been made in attempts to explain these observations, but have limited applicability beyond low photon numbers \cite{Boissonneault2009,Malekakhlagh2020,Petrescu2020}.

In this paper, we go beyond perturbative treatments by numerically investigating the full dynamics of a strongly driven transmon-resonator system. At distinct resonator populations, we find clear signatures of transmon ionization where transmon states above the Josephson junction potential are occupied~\cite{Verney_2019,Lescanne_2019}. Because these states are not strongly influenced by the Josephson potential, they are well described by charge states. Consequently, for states above the transmon well the transmon-resonator coupling appears longitudinal and the system dynamics is consequently modified. 

Accurately simulating the dynamics of transmon ionization requires describing the system's density matrix on a truncated Hilbert space of very large dimension, something that is made possible here by the use of large-scale computational accelerators known as Tensor Processing Units (TPUs), introduced below. While other studies were limited to steady-state calculations with strongly-detuned drives~\cite{Verney_2019}, the computational power of TPUs allows us to simulate the full time dependence with drives that are resonant with the resonator, as is relevant for qubit measurement. Accounting for tens of transmon levels and hundreds of resonator states, we moreover see signatures of the high-power readout~\cite{Reed2010,Bishop2010,Boissonneault2010}.
We interpret these numerical results using a semiclassical model capturing the nonlinear response of the driven system. Using this model, we identify parameter regimes where ionization is expected to occur at sufficiently small photon number to affect dispersive readout, observations which are in qualitative agreement with experiments~\cite{Walter2017}.

This paper is organized as follows. In \cref{section:MasterEquationTPUS} we introduce the model, provide details on its  TPU implementation, and present the numerical results on the dynamics of transmon ionization. Next, in \cref{section:Semiclassical} we formulate a semiclassical theory which allows us to interpret the dynamics of the coupled transmon-resonator system as governed by transitions between qubit-state-dependent effective resonators 
(\cref{section:identification}) obeying nonlinear equations of motion (\cref{section:EOMS}). In \cref{section:WalterEtAl}, we choose system parameters illustrating how ionization can reduce readout fidelity even at the low drive powers that are typical of dispersive readout in circuit QED. We summarize our findings in \cref{section:Conclusion}.

\section{Master Equation Simulations with Tensor Processing Units (TPUs)}
\label{section:MasterEquationTPUS}

\subsection{Model}
\label{subsn:MEModel}

\begin{figure}
    \centering
    \includegraphics[width=1\columnwidth]{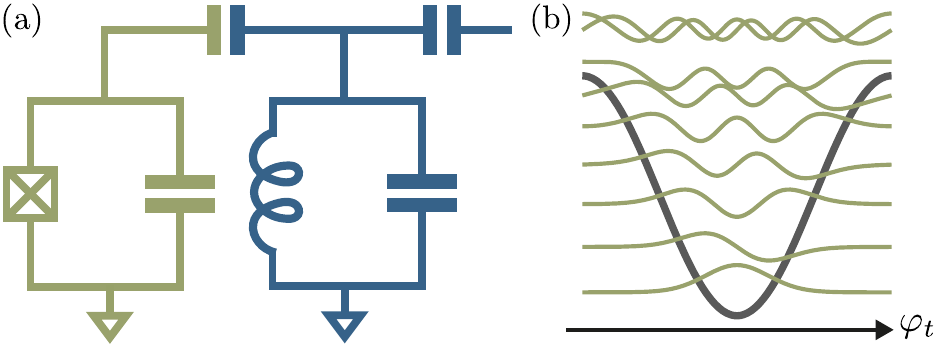}
    \caption{(a) Transmon (green) capacitively coupled to a driven resonator (blue). (b) Cosine potential of the transmon with its first 9 eigenstates. For our choice of parameters, the last three states are ionized states lying above the top of the potential.
    }
    \label{fig:TransmonResonator}
\end{figure}

We consider a transmon capacitively coupled to a resonator, see \cref{fig:TransmonResonator}(a). In the presence of a drive of amplitude $\mathcal{E}$ and frequency $\wdrive$ on the resonator, the Hamiltonian of the system takes the usual form ($\hbar = 1$) \cite{Blais2021} 
\begin{equation}
    \begin{split}\label{eqn:DrivenSystemHamiltonian}
        \hat{H}(t) & = 4E_c \hat{n}_t^2 -E_J\cos\hat{\varphi}_t + \omega_r \hat{a}^\dagger \hat{a}\\ 
        &- ig\hat{n}_t(\hat{a} - \hat{a}^\dagger) - i \mathcal{E}( \hat{a}-\hat{a}^\dagger)\sin(\omega_d t). 
    \end{split}
\end{equation}
The first two terms correspond to the free transmon Hamiltonian with charging energy $E_c$, Josephson energy $E_J$, charge 
operator $\hat{n}_t$ and phase operator $\hat{\varphi}_t$. We denote the eigenenergies and eigenstates of the free transmon Hamiltonian $E_i$ and $\ket{i}$, respectively. The first two of those eigenstates, labelled $\lbrace |0\rangle, |1\rangle \rbrace$, span the computational basis of the qubit. Of the higher excited states, a number $\sim 2E_J/\omega_p$ are bound states lying within the cosine potential well illustrated in \cref{fig:TransmonResonator}(b). Here, $\omega_p = \sqrt{8E_CE_J}$ is the plasma frequency which is approximately the transmon's 0-1 transition frequency~\cite{Koch_2007_Transmon}. Moreover, we label $\ket{n}$ the eigenstates of the free resonator Hamiltonian of frequency $\omega_r$, corresponding to the bosonic annihilation operator $\hat{a}$. The transmon-resonator coupling of amplitude $g$ in the second line of \cref{eqn:DrivenSystemHamiltonian} includes fast-rotating terms that are important to capture the contribution of high-energy states~\cite{Sank2016}. 
In the absence of the drive, the dressed energies and states of the coupled system are denoted $E_{\overline{i,n}}$ and $\ket{\overline{i,n}}$.

Including cavity loss at a rate $\kappa$, the driven transmon-resonator system is described by the usual Lindblad master equation \cite{Blais2021}
\begin{equation}
    \dot{\hat{\rho}} = -i[\hat{H}(t),\hat{\rho}] + \kappa  \mathcal{D}[\hat{a}]\hat{\rho},
    \label{eqn:MasterEquation}
\end{equation}
with the dissipator
\begin{equation}
    \mathcal{D}[\hat O]\bullet = \hat O \bullet \hat O^\dag - \frac{1}{2} \left\{\hat O^\dag \hat O, \bullet \right\}.
\end{equation}

Because we are interested in capturing the dynamics of the system in the presence of a large amplitude drive on the resonator, leading to several hundred of photons and highly excited states of the transmon, we keep up to $32$ states in the transmon and $1,024$ states in the resonator, corresponding to a total Hilbert space dimension $2^{15} =$ 32,768. That is, the joint density matrix $\hat{\rho}$ for the transmon-resonator system is a Hermitian matrix of size $2^{15}\times 2^{15}$, which thus contains $2^{30} =$ 1,073,741,824 time-dependent complex coefficients. Furthermore, given that the unbound transmon states are approximately eigenstates of the charge operator, their eigenvalues  increase quadratically rather than linearly as is the case for the bound states. This increases the complexity of numerical simulations, as larger eigenvalues require smaller integration step sizes for convergence. More details are provided in \cref{section:NumericalImplementation}.

In order to perform these challenging numerical simulations, we resort to Tensor Processing Units (TPUs). Google's TPUs are application specific integrated circuits designed exclusively to accelerate large-scale machine learning workloads \cite{TPUinfo}. Recently, they have been repurposed for other high-performance computational tasks \cite{TPUFFT1, TPUFFT2, huot2019highresolution, Belletti-Anderson2020,  Lu-Ma2020, Wang-Anderson2021, tpu_algebra, tpu_qchem, tpu_floquet, tpu_qphys, tpu_Z2field, tpu_circuit, tpu_DMRG}, including simulations of quantum systems in large Hilbert spaces \cite{tpu_floquet, tpu_qphys, tpu_Z2field, tpu_circuit, tpu_DMRG}. Thanks to their (i)~matrix multiply units capable of accelerating matrix multiplication, (ii)~large amounts of high-bandwidth memory and (iii)~fast inter-core interconnects directly connecting up to thousands of cores, TPUs are particularly fast at performing large-scale dense linear algebra operations, which are required e.g.~to numerically integrate the above Lindblad master equation. More details on TPUs can be found in \cref{section:TPUdetails}.

For a given drive amplitude $\mathcal{E}$, simulations begin with the system initialized in either the dressed ground $\ket{\overline{0,0}}$ or excited $\ket{\overline{1,0}}$ states. To reduce the simulation time while capturing the transient dynamics including the ionization of the transmon, we simulate the evolution over a time at least $\kappa^{-1}$. After each period of the drive, the reduced transmon, $\hat{\rho}_t$, and resonator, $\hat{\rho}_r$, density matrices are recorded. 

Except otherwise stated, in all of our simulation as well as in the semiclassical model discussed in \cref{section:Semiclassical} we use the parameters $E_J/E_C = 50$, $E_C/h =280$ MHz
corresponding to approximately $6$ bound transmon states within the cosine potential, see \cref{fig:TransmonResonator}. The resonator frequency is $\omega_r/2\pi = 7.5$ GHz and $g/2\pi = 250$ MHz. A resonator loss rate of $\kappa/2\pi = 20$ MHz is chosen to ensure fast resonator dynamics and the drive amplitude takes values in the range $\mathcal{E}/2\pi \in [0,450]$~MHz. We fix the drive frequency to the bare resonator frequency, $\omega_d =\omega_r$. The above parameters place the system in the dispersive regime with $\chi/\kappa = 0.28$~\footnote{The dispersive shift is defined here as $\chi = (E_{11}-E_{10}-E_{01}+E_{00})/2$.} when the resonator population is much smaller than the critical photon number $n_{\textrm{crit}} = (\Delta/2g')^2 \approx 15$, with $g' = (E_J/32E_C)^{1/4}g$ \cite{Blais2021}.

\subsection{Numerical Results}
\label{subsn:NumericalResults}

\begin{figure*}
    \centering
    \includegraphics[width=\linewidth]{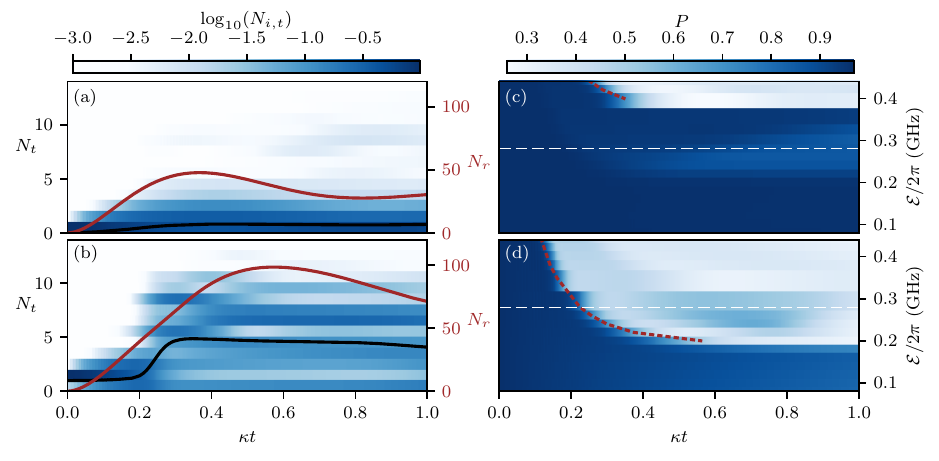}
    \caption{(a,b) Transmon and photon population as a function of time with $\mathcal{E}/2\pi = 0.28$ GHz for (a) the ground state and (b) the excited state. The black line indicate the average transmon population $N_t$ (left axis), and the solid red line depicts the resonator average population $N_r$ (right axis). The full distribution $N_{i,t}$ of the transmon population for each transmon levels  $\ket i$ is also plotted, with the color bar above providing the scale. (c-d) Purity of the reduced transmon density matrix as a function of the drive amplitude and time for the transmon initialized (c) in the ground state and (d) in the excited state. The dashed red lines indicate when the resonator reaches a population of 105 and 42 photons for the ground and excited states, respectively. The white dashed lines indicate drive amplitude $\mathcal{E}/2\pi = 0.28$ GHz of panels (a,b). Because the drive amplitude is not sufficiently large to reach a population of $N_r = 105$ photons when the qubit is the ground state, ionization is not observed in panel (a).
    }
    \label{fig:PuritiesDrivevsTimeGE}
\end{figure*}

\begin{figure}
    \centering
    \includegraphics[width=\columnwidth]{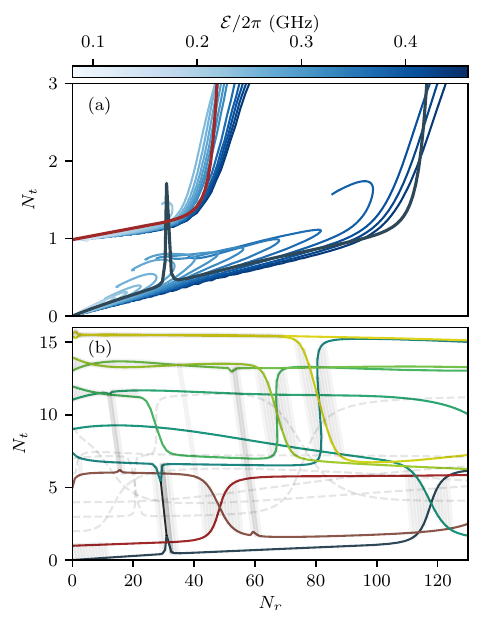}
    \caption{(a) Parametric plot of the average transmon population versus the average resonator population for an  evolution time of $\kappa^{-1}$. Each line represents a unique simulation with a different drive amplitude $\mathcal{E}$ and starting either in the ground or excited state. (b) Average transmon population as a function of the bare resonator population for each resonator branch $\{i\}$. The nearly vertical dark lines represent the overlap $C_i(n)$ between the different branches, darker lines indicating stronger overlap. 
    } 
    \label{fig:AllbranchesFadedInteractions}
\end{figure}

\cref{fig:PuritiesDrivevsTimeGE}(a,b) illustrate the average photon number ${N}_r = \langle \hat{a}^\dagger \hat{a} \rangle$ (red lines) and the average transmon population ${N}_t =  \sum_i i \langle i | \hat{\rho}_t | i \rangle$ (black line) as a function of time for the drive amplitude  $\mathcal{E}/2\pi = 280$~MHz. The system is initialized in the dressed ground state $\ket{\overline{0,0}}$ in panel (a) and in the dressed excited state $\ket{\overline{1,0}}$ in panel (b). Also shown is the instantaneous distribution $N_{i,t} = \langle i | \hat{\rho}_t | i \rangle$ of the transmon states (color scale). The difference between the two initial states is striking. For this drive amplitude, when starting in the ground state the transmon leaks out of its initial state but the distribution largely remains confined within the cosine potential of the Josephson junction. In contrast, when the transmon is initialized in its excited state we observe a sudden jump of the average transmon population with a distribution extending well above the top of the cosine potential. This is a clear illustration of the ionization of the transmon. For the drive amplitude used here, the simulated measurement is far from QND and the dynamics therefore not well described by a dispersive Hamiltonian.

As illustrated in \cref{fig:PuritiesDrivevsTimeGE}(c,d), transmon ionization is accompanied by a sudden drop in the purity $P = \textrm{Tr}[\hat{\rho}_t^2]$ of the reduced transmon density matrix. The rapid decline of the purity is observed at specific resonator photon populations indicated by the dashed red line: $N_r = 105$ when initialized in the ground state in panel (c) and $N_r = 42$ for the excited state in panel (d). The dashed red lines terminate whenever the drive is too weak for the resonator to reach those populations. That transmon ionization  occurs at specific photon populations suggests that the phenomenon is due to resonances, which we discuss in more detail below. These observations are compatible with Ref.~\cite{Verney_2019} where steady-state calculations with an off-resonant drive also showed drops in transmon purity in steady-state numerical calculations. Resonances at large photon numbers have also been observed in Ref.~\cite{Sank2016}.

As a further illustration of transmon ionization at distinct photon numbers, we plot in \cref{fig:AllbranchesFadedInteractions}(a) the transmon population $N_t$ as a function of the resonator population $N_r$, both values taken from time traces such as shown in \cref{fig:PuritiesDrivevsTimeGE}(a,b). The different curves correspond to different drive amplitudes, and the initial state of the transmon is identifiable from the starting transmon population. Remarkably, the responses essentially collapse to single curves for each of the initial states. Below the ionizing photon population, the transmon population remains close to the computational manifold and the resonator population exhibit transient oscillations due to the drive being off-resonant with the Lamb-shifted resonator frequency. Together with Purcell decay, these transient oscillations are responsible for the features observed at small $N_r$. Above the ionizing photon populations, the transmon population rapidly increases. This coincides with the sudden drop of purity observed in \cref{fig:PuritiesDrivevsTimeGE}(c,d).

\begin{figure*}
    \centering
    \includegraphics[width=2\columnwidth]{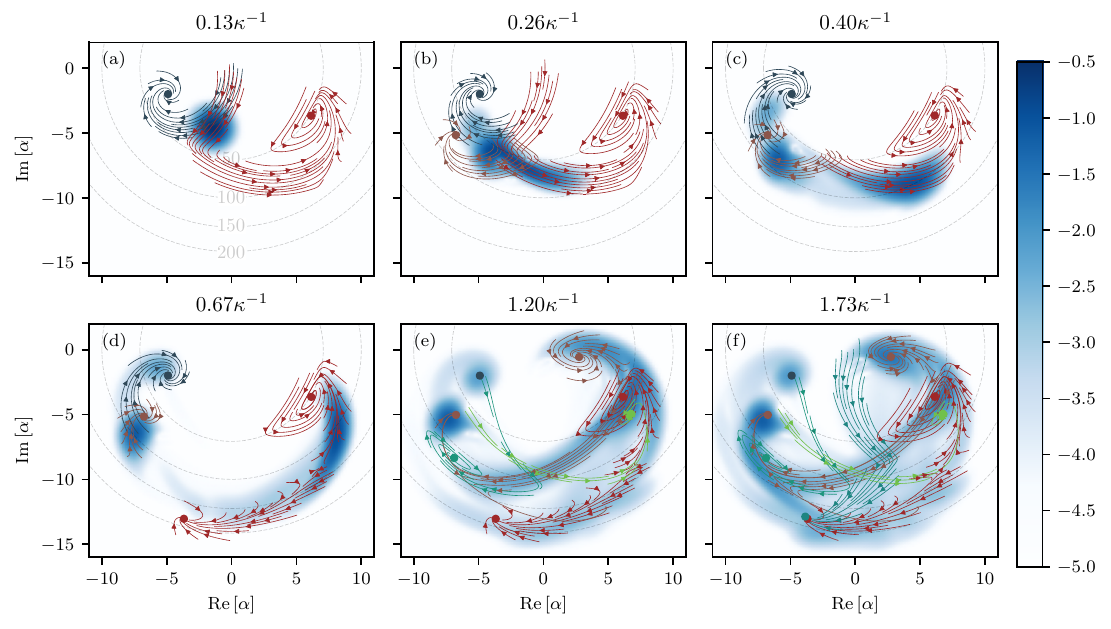}
    \caption{Wigner function of the resonator at different times for a drive amplitude $\mathcal{E}/2\pi = 0.28$ GHz and the transmon initialized in its dressed excited state. For small features to be observable, the logarithm of the absolute value of the Wigner function is plotted. Overlaid are the semiclassical flows of the relevant branches following the color convention of \cref{fig:AllbranchesFadedInteractions}(b). The dashed circles are guide to the eyes at constant photon numbers (50, 100, 150, and 200).
   }
    \label{fig:WignersWithTrajectories}
\end{figure*}

Focusing now on the dynamics of the resonator, \cref{fig:WignersWithTrajectories} shows the Wigner function of the resonator at different times with the transmon starting in its excited state and with the same drive amplitude $\mathcal{E}/2\pi = 0.28$ GHz as used in \cref{fig:PuritiesDrivevsTimeGE}(a,b). For small features to appear more clearly, we plot the logarithm of the absolute value of the Wigner function. As expected, the cavity is initially well described by the single coherent states $\ket{\alpha_1}$ associated to the transmon's first excited state, see panel (a). As the transmon population in other levels increases, additional features emerge corresponding to the coherent states associated to the now occupied transmon levels.  Additionally, ``bananization" caused by transmon-induced nonlinearities become apparent \cite{Boutin2017,Sivak2019}. The dashed lines correspond to fixed photon numbers and are used as guide to the eye. The full colored lines overlaying the Wigner functions are obtained from a semiclassical approximation which we now introduce.

\section{State Identification and Semiclassical Interpretation}
\label{section:Semiclassical}

The TPU-based, large-scale numerical simulations presented in the previous section clearly illustrate the breakdown of the dispersive approximation. Notably, we observe at specific resonator photon populations a sudden jump in transmon population above the cosine potential well which is associated with a sharp drop in transmon purity. This results in complex dynamics of the resonator field, as illustrated by its Wigner function. In this section, we develop a semiclassical model to understand the main features of these observations.

\subsection{Identification of dressed states and resonator branches}
\label{section:identification}

Our semiclassical approach is based on the dressed states $\ket{\overline{i,n}}$ and energies $E_{\overline{i,n}}$ of the transmon-resonator Hamiltonian \cref{eqn:DrivenSystemHamiltonian} in the absence of the drive. More precisely, at arbitrary photon number we identify the dressed states which are closest to transmon eigenstates. At low photon number this identification is simple in the dispersive regime. However, as the photon population approaches and even exceeds $\ncrit$, the dressed states are highly entangled and identification becomes difficult.

Building on Ref.~\cite{Boissonneault2010}, our approach relies on first identifying the eigenstates of the Hamiltonian of \cref{eqn:DrivenSystemHamiltonian} for $\mathcal{E} = 0$ obtained from numerical diagonalization with the largest overlap with $|{i,0\rangle}$, the bare transmon states at zero photon population. Then, given an identified state $\ket{\overline{i,n}}$ for $n \geq 0$, the next state $\ket{\overline{i,n+1}}$ is chosen from the subset of remaining eigenstates $\lbrace |\lambda\rangle \rbrace$ such that the overlap
\begin{equation}
    C_i(n) = \left| \frac{\bra{\lambda}\hat{a}^\dagger\ket{\overline{i,n}} }{\bra{\overline{i,n}}\hat{a} \hat{a}^\dagger \ket{\overline{i,n}}} \right|^2
    \label{eqn:OverlapIdentifyladders}
\end{equation}
with the state $\ket{\overline{i,n}}$ plus one additional resonator excitation is maximum.
Following this procedure recursively we obtain a set of states $\lbrace \ket{\overline{i,n}} \rbrace$ where the bare transmon label $i$ is fixed and $n$ spans a desired range of resonator population. We refer to each such set of states as a branch $\lbrace i \rbrace$ of the resonator. In \cref{fig:AllbranchesFadedInteractions}(b), we plot the average transmon population of the first 16 of these resonator branches as a function of photon number. The full coloured lines are branches which play an important role in the understanding of the numerical results. Branches playing a more minor role for our particular choice of parameters are illustrated as dashed gray lines.

At finite transmon-resonator coupling, and because we include non-RWA terms in the system Hamiltonian, the different branches have a non-zero overlap $C_i(n)$. To illustrate this, in \cref{fig:AllbranchesFadedInteractions}(b) we plot with dark and nearly vertical lines the overlap $C_i(n)$ between branches whenever it rises above $0.01$. Darker lines indicate a stronger overlap between the states. At small photon number, the overlap between $\{0\}$ and $\{1\}$ corresponds to Purcell decay which is seen to decrease with increasing photon number~\cite{Sete2014}. More importantly, we note a strong overlap between the branches $\lbrace 0 \rbrace$ and $\lbrace 9 \rbrace$ for $N_r \sim 110$, and between $\lbrace 1 \rbrace$ and $\lbrace 5 \rbrace$ for $N_r \sim 50$, in agreement with the photon numbers at which sudden drops of transmon purity is observed in \cref{fig:PuritiesDrivevsTimeGE}(c,d). To further emphasize the link between the large overlaps and transmon ionization, we reproduce as solid blue and red lines the first two branches $\lbrace 0 \rbrace$ and $\lbrace 1 \rbrace$ of \cref{fig:AllbranchesFadedInteractions}(b) together with the numerical data in \cref{fig:AllbranchesFadedInteractions}(a). The sharp elbows, which are indicative that transmon ionization results from a resonance, align well between the semiclassical results and the numerical data. 

A strong overlap is also observed between $\{0\}$ and $\{8\}$ close to $N_r = 30$, see \cref{fig:AllbranchesFadedInteractions}(a,b). In contrast to what is observed at $N_r \sim 50$ and 110, this sharper resonance results in very little population transfer. This can be understood from the framework of Landau-Zener transitions. In our semiclassical picture, as the resonator rings up towards its transmon-state-dependent steady state, the photon number is swept at a rate related to $\kappa$ and to the drive amplitude. A large drive amplitude leads to a rapid sweep through the narrow feature at $N_r \sim 30$ resulting in a diabatic passage with little leakage, see the darker blue numerical lines in \cref{fig:AllbranchesFadedInteractions}(b). Imperfect diabatic transitions at $N_r \sim 30$ can be observed for smaller drive amplitudes. On the other hand, for our choice of parameters, the features at $N_r \sim 50$ and 110 are broader and the sweep rate therefore comparatively slower. This results in a non-diabatic process leading to leakage out of the transmon computational subspace and to the observed drop in purity. This interpretation is further confirmed in \cref{sec:LZ} where numerical experiments with a different drive frequency and loss rates are presented.

\subsection{Semiclassical dynamics}
\label{section:EOMS}
\begin{figure}
    \centering
    \includegraphics[width=\columnwidth]{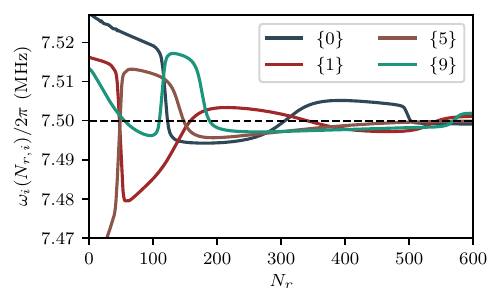}
    \caption{Effective frequency $\omega_i(N_r)$ of resonator branches as a function of the resonator population. The frequency has been smoothed using a Savitzky–Golay filter to remove discontinuities caused by strong interactions with other branches. The dashed line indicates the bare cavity frequency $\omega_r/2\pi = 7.5$ GHz.
    }
    \label{fig:resonatorFreqvsPhotNum}
\end{figure}

Following Ref.~\cite{Boissonneault2010}, to each branch $\lbrace i \rbrace $ we assign an effective oscillator of photon-number dependent frequency
\begin{equation}
    \omega_i(n) = E_{\overline{i,n+1}} - E_{\overline{i,n}}.
    \label{eqn:omegain}
\end{equation}
As seen in \cref{fig:resonatorFreqvsPhotNum} where we plot  $\omega_i(n)$ versus $n$, \cref{eqn:omegain} accounts for the photon-number dependence of the resonator frequency, including that the resonator responds at its bare frequency at large photon numbers. Assuming that the transmon remains in a given branch $\lbrace i \rbrace$, the dynamics of this effective oscillator approximately obeys the classical equation of motion of a driven damped oscillator
\begin{equation}
    \dot{\alpha_i} = -i[\omega_i(|\alpha_i|^2) -\omega_d]\alpha_i -i\mathcal{E}/2 -\kappa\alpha_i/2 ,
    \label{eqn:DifferentialEquationNonlinearresonator}
\end{equation}
where nonlinear effects are encapsulated in the dependence of the branch frequency on photon number $\omega_i(|\alpha_i|^2)$ according to \cref{eqn:omegain}. Because the quantity $|\alpha_i|^2$ takes arbitrary real values, we generalize  \cref{eqn:omegain} from discrete values $n$ to a continuous function. To do this, we smooth $\omega_i(n)$ with a first-order Savitzky–Golay filter and linearly interpolate between each $n$. Doing this additionally removes large discontinuities in the photon-number dependence of the effective frequencies $\omega_i(n)$ caused by strong interactions with other branches. This semiclassical approximation is expected to accurately describe the system for large photon number $|\alpha_i|^2 \gg 1$.

Numerically integrating \cref{eqn:DifferentialEquationNonlinearresonator}, we plot in \cref{fig:WignersWithTrajectories}(a) the time dependence of $\alpha_1(t)$ (red lines) together with a snapshot of the Wigner function at time $t=0.13\kappa^{-1}$ assuming that the effective oscillator starts in vacuum. The arrows indicate the flow of time and the different lines are obtained from \cref{eqn:DifferentialEquationNonlinearresonator} for a set of initial conditions in an area close to vacuum representing zero point fluctuations. The red dot corresponds to the steady-state value $\alpha_1^s$. 

To account for transitions between transmon states, the evolution of $\alpha_i(t)$ associated to other branches is illustrated whenever the semiclassical evolution reaches photon numbers at which we anticipate non-negligible rates for transitions into those branches according to the overlaps observed in \cref{fig:AllbranchesFadedInteractions}(b). To distinguish between the different branches we use the color scheme of  \cref{fig:AllbranchesFadedInteractions}(b). For example, the blue lines in \cref{fig:WignersWithTrajectories}(a-d) correspond to branch $\lbrace 0 \rbrace$ which appears as a result of Purcell decay. In the same way, starting in panel (b) we include the flow of $\alpha_5(t)$ associated to branch $\lbrace 5 \rbrace$  (brown lines) which has a strong overlap with $\lbrace 1 \rbrace$ when $N_r \sim 50$. Following the evolution of the Wigner function from panel to panel, it is possible to see a feature following the flow of $\alpha_5(t)$ and settling at the expected steady-state value $\alpha_5^s$ in the last panel (brown dot). Applying this procedure whenever the resonator number is such that an occupied branch has a strong overlap $C_i(n)$ with another branch, the vast majority of the Wigner function transient behaviour can be understood. In particular, starting in panel (d) we see the bistable behavior of $\{\alpha_1\}$ in both the semiclassical results (red lines) and the numerical data.

\begin{figure}
    \centering
    \includegraphics[width=\columnwidth]{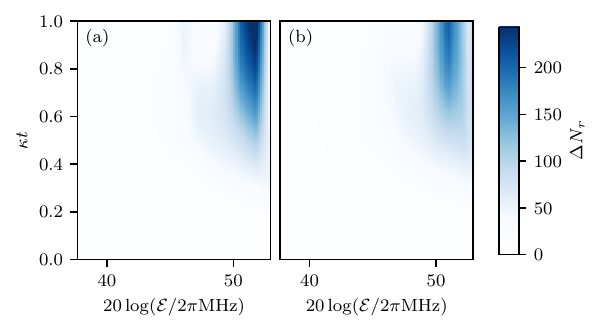}
    \caption{Difference in resonator population $\Delta N_r$ given that the transmon is initialized in its ground or excited state as a function of the drive amplitudes and time as obtained from (a) the semiclassical model and  (b) the numerical TPU data.}
    \label{fig:HighPowerReadoutSnapshots}
\end{figure}

Rather than focusing on the steady-state response, in \cref{fig:HighPowerReadoutSnapshots}(a) we plot, for different drive amplitudes and as a function of time, the difference in the resonator populations $\Delta N_r$ obtained by solving the semiclassical expression \cref{eqn:DifferentialEquationNonlinearresonator} given that the transmon is initialized in its ground or excited state. As expected from \cref{fig:resonatorFreqvsPhotNum}, at some threshold power the photon number rapidly increase if the qubit is initially in its excited state while the increase is not as pronounced for ground state. This results in the observed large $\Delta N_r$. Panel (b) shows the same quantity obtained from numerical simulations. Given the large photon numbers that are involved, those simulations are particularly demanding and required Hilbert space sizes up to $2^{15}$. The agreement between the full numerical simulations and the simple semiclassical model is remarkable. In particular, in both approaches we observe that the photon number difference $\Delta N_r$ goes to zero at the strongest drive amplitudes. This is expected from  \cref{fig:resonatorFreqvsPhotNum} where at very large photon numbers the frequencies $\omega_i(|\alpha_i|^2)$ eventually collapse to the bare value $\omega_r$. The different drive amplitude at which this collapse occurs depend on the initial transmon state and the resulting large $\Delta N_r$ is exploited in the high-power qubit readout~\cite{Reed2010, Boissonneault2010,Bishop2010}. 

A possible interpretation for the observed large response at the bare resonator frequency is that once the transmon is ionized, mostly charge-like states are occupied. Because these states couple longitudinally to the resonator, they do not lead to a resonator frequency pull~\cite{Lescanne_2019}. However, numerical results show that even for the highest drive powers a significant distribution of states inside the well remain populated (not shown). As a result, the collapse of the resonator to its bare frequency cannot be explained as resulting alone from the longitudinal-type coupling of the unbound states. We leave a more detailed analysis of this effect to future work~\cite{Cohen2022}.

\subsection{Resonances at low photon number}
\label{section:WalterEtAl}

\begin{figure}
    \centering
    \includegraphics[width=\columnwidth]{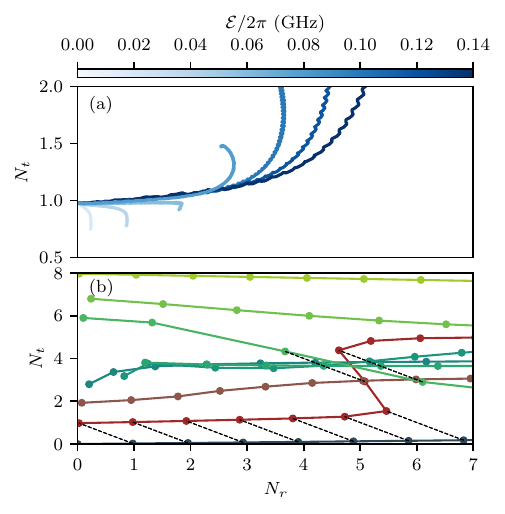}
    \caption{(a) Parametric plot of the average transmon population versus the average resonator population for an evolution time of $48$ ns. Each line represents a unique simulation with a different drive amplitude $\mathcal{E}$ and starting in the excited state. Because of the small Hilbert space size, these results are obtained using CPUs. (b) Average transmon population as a function of the bare resonator population for each resonator branch $\{i\}$. The dashed lines indicate the overlap between different branches when $C_i(n)>0.01$.}
    \label{fig:WalterBranchPlot}
\end{figure}

For the parameters used above, the first resonance leading to non-QNDness occurs at a relatively large photon numbers of $\sim 30$, followed by resonances at even larger photon numbers. Such resonances can, however, occur at much lower photon numbers. To demonstrate this, we now use parameters based on the experiment of Ref.~\cite{Walter2017}: $E_C/h = 314$~MHz, $E_J/E_C = 55.47$, $g/2\pi= 211$~MHz, $w_r/2\pi=4.804$~GHz, corresponding to $\ncrit \approx 10$. The cavity damping rate is set to $\kappa/2\pi = 40$~MHz. The drive amplitude is varied in the range $\mathcal{E}/2\pi \in [0.02,0.14]$~MHz and the evolution time is set to 48~ns corresponding to the smallest measurement time used in Ref.~\cite{Walter2017}. We note that these are bare system parameters chosen such as to approach the dressed parameters reported in Ref.~\cite{Walter2017}.

In a similar fashion to \cref{fig:AllbranchesFadedInteractions}(a), in \cref{fig:WalterBranchPlot}(a) we parametrically plot the transmon population $N_t$ against the resonator population $N_r$ for the transmon initialized in the excited state. For this choice of parameters and integration time, Purcell decay is apparent for the three lowest drive amplitudes (see the sharp decrease of $N_t$). For larger measurement amplitudes corresponding to an average resonator population of approximately 2.5 photons, upward transition of the transmon population is clearly observed. Interestingly, this corresponds to the average photon number at which a decrease in measurement fidelity is observed in Ref.~\cite{Walter2017}, see Fig.~3(b) of that reference.

To understand the origin of this population leakage under measurement, we show in \cref{fig:WalterBranchPlot}(b) the average photon population for the resonator branches $\{i\}$ as obtained from our semiclassical model. At $N_r\approx 5$, the transmon population rapidly rises for branch $\lbrace 1 \rbrace$ associated with the transmon's first excited state (red line). In the numerical simulation shown in \cref{fig:WalterBranchPlot}(a) as well as in experiments, the resonator field is in a coherent state. As a result, because of the $\sqrt{n}$ fluctuations of coherent states, the resonance at $N_r\approx 5$ already plays a role at $N_r\approx2.5$ leading to the observed non-QNDness. While care must be taken when using a semiclassical model at such small photon population, these observations are suggestive that the resonances that are observed here could play a role in the drop of measurement fidelity and QNDness that is experimentally observed at low photon number.

\section{Conclusion}
\label{section:Conclusion}

We leverage the computational power of TPUs to perform large Hilbert-space size time-dependent master equation simulations of transmon readout in circuit QED. From these simulations, we identify resonances occurring at distinct resonator photon number where the transmon population escapes to states above the Josephson cosine potential well. To interpret these results, we develop a semiclassical model capturing the nonlinear transmon-state dependent change of the resonator frequency with photon number. In particular, this model correctly captures which states play a significant role in transmon ionization. Using a different set of parameters, we show that these resonances can occur at small photon population. These results suggest that the non-QND nature of dispersive readout which is experimentally observed at small drive amplitudes could be due these resonances. The semiclassical model developed here can be used as a tool to avoid these spurious effects when optimizing device parameters for readout.

\section{Acknowledgements}
This work was undertaken in part thanks to funding from NSERC, the Canada First Research Excellence Fund, the Minist\`ere de l’\'Economie et de l’Innovation du Qu\'ebec, the U.S. Army Research Office under Grant No. W911NF-18-1-0411, and the U.S. Department of Energy, Office of Science, National Quantum Information Science Research Centers and Quantum Systems Accelerator.
This research was also supported by Cloud TPUs from Google's TPU Research Cloud (TRC). 
GV is a Distinguished Invited Professor at the Institute of Photonic Sciences (ICFO), and a Distinguished Visiting Research Chair at Perimeter Institute. Research at Perimeter Institute is supported by the Government of Canada through the Department of Innovation, Science and Economic Development and by the Province of Ontario through the Ministry of Research, Innovation and Science.

\bibliography{references}

\appendix

\section{Numerical Implementation}
\label{section:NumericalImplementation}

To simulate the master equation in \cref{eqn:MasterEquation}, we move to a rotating frame defined by
\begin{equation}
    U_{\textrm{rf}}(t) = \exp(i\omega_d a^\dagger a t).
\end{equation}
In that frame, the system Hamiltonian takes the form
\begin{equation}
\begin{aligned}
\hat{H}_I(t) 
&= \hat{U}_{\textrm{rf}}(t)\hat{H}(t)U^\dagger_{\textrm{rf}}(t) - i U_{\textrm{rf}}(t) \dot U^\dag_{\textrm{rf}}(t) ,\\
&= \hat{H}_{I0} + \hat{X}_{I}(t),
\end{aligned}
\end{equation}
with
\begin{equation}
    \hat{H}_{I0} = 4E_c \hat{n}_t^2 -E_J\cos(\hat{\varphi}_t) + (\omega_r-\omega_d) \hat{a}^\dagger \hat{a} -\frac{\mathcal{E}}{2} (\hat{a}^\dagger+\hat{a}),
\end{equation}
and
\begin{equation}
    \hat{X}_I(t) = - ig\hat{n}_t(\hat{a}^\dagger e^{i\omega_d t} - \hat{a}e^{-i\omega_d t}) +\frac{\mathcal{E}}{2} (\hat{a}^\dagger e^{2i\omega_d t}+\hat{a}e^{-2i\omega_d t}).
\end{equation}
Moreover, in this rotating frame the master equation reads 
\begin{equation}
    \dot{\hat{\rho}}_I(t)  =  -i[H_I(t),\hat{\rho}_I] + \kappa \mathcal{D}[\hat{a}]\hat{\rho}_I \equiv \hat{\mathcal{L}}\hat{\rho}_I,
\end{equation}
with ${\hat{\rho}}_I(t) = \hat{U}_{\textrm{rf}}^\dagger(t) {\hat{\rho}}(t) \hat{U}_{\textrm{rf}}(t)$ and 
where the dissipator $\mathcal{D}[\hat{a}]\bullet$ is unaffected by the transformation. 

To solve this master equation on TPUs we approximate the action of the time-ordered Lindbladian exponential $\mathcal{T}\exp\big(\int_0^t \hat{\mathcal{L}}(t')dt'\big)\bullet$. To do this, we first solve for the roots $\lbrace z_i \rbrace$ of the Taylor approximation of the exponential to $n$-th order and rearrange the terms
\begin{equation}
    e^x \approx \sum_{i=0}^n \frac{x^i}{i!} = \frac{1}{n!}\prod_{i=1}^n(x-z_i) = \prod_{i=1}^n (1-x/z_i),
\end{equation}
where $\prod_i z_i = n!$. We then evolve the system by time $\delta t$ by iterating through the $n$ roots:
\begin{equation} \label{eq:exp_evolve}
\begin{aligned}
\hat{\rho}(t+\delta t) &= \hat{\rho}_n(t),~~~\hat{\rho}_{i=0}(t) = \hat{\rho}(t)\\
    \hat{\rho}_{i}(t) &= \left[1- (\delta t/z_i)\overline{\hat{\mathcal{L}}(t,t+\delta t)}\right]\hat{\rho}_{i-1}(t), \\
\end{aligned}
\end{equation}
where $ \overline{\hat{\mathcal{L}}(t,t+\delta t)}\bullet$ is calculated using a second-order Magnus expansion. The order $n$ and step size $\delta t$ required for convergence depend on the number of transmon and resonator states, with values of $n=8-15$ and $1/(\omega_d\delta t) \approx 50-100$ found to be optimal for our choice of parameters.

The Hilbert space of the cavity was adapted throughout the evolution. At each step the quantity $E = |1-\Tr \lbrace \rho(N\delta t)[\hat{a}, \hat{a}^\dagger] \rbrace|$ was calculated to ensure minimal occupation of the highest cavity excited states. The cavity size was doubled if $E > 10^{-6}$, with the previous step recalculated with the larger Hilbert space size once this condition was met.

\section{Tensor Processing Units}
\label{section:TPUdetails}

Google's TPUs are application specific integrated circuits (ASICs) originally designed to accelerate and scale up machine learning workloads~\cite{TPUinfo}. By leveraging the JAX library~\cite{jax}, it is possible to repurpose TPUs to also accelerate other large-scale computational tasks~\cite{TPUFFT1, TPUFFT2, huot2019highresolution, Belletti-Anderson2020,  Lu-Ma2020, Wang-Anderson2021, tpu_algebra, tpu_qchem, tpu_floquet, tpu_qphys, tpu_Z2field, tpu_circuit, tpu_DMRG}. For instance, in Refs.~\cite{tpu_floquet, tpu_qphys, tpu_Z2field, tpu_circuit} TPUs are used to simulate the wavefunction of up to 36-40 (two-level) qubits. In this work we have used the power of TPUs to simulate instead the time evolution of the joint density matrix of a transmon-resonator system, as given by a Lindblad master equation. 

We employed TPUs of third generation, denoted v3. Each single TPU v3 core is equipped with two matrix multiply units (MXUs) to formidably accelerate matrix-matrix multiplication (matmul), resulting in about 10~TFLOPS of measured single-core matmul performance in single precision. 

The smallest available TPU configuration consists of 8 TPU v3 cores with a total of 128 GB of dedicated high-bandwidth memory (HBM), controlled by a single host with 48 CPU cores. The largest configuration is a pod with 2048 TPU v3 cores and 32 TB of HBM, controlled by 256 hosts. Given a choice of configuration, the available TPU cores are directly connected to nearest neighbors in a 2D torus network through fast inter-core interconnects (ICIs). The ICIs play an essential role in the TPUs ability to maintain high performance when distributing matmuls and other dense linear algebra operations over all available TPU cores. In this work we used the JAX library~\cite{jax} to write single program multiple data (SPMD) code and executed it on configurations made of multiple TPU cores. Specifically, for the largest density matrix under consideration, of size $2^{15} \times 2^{15}$, we used 128 TPU v3 cores. The density matrix was distributed over all available cores and updated according to the different terms in the Lindblad operator, with a typical update $\hat{\rho}(t) \rightarrow \hat{\rho}(t+\delta t)$ for $n=15$ in \cref{eq:exp_evolve} taking on the order of seconds.

\begin{figure}
    \centering
    \includegraphics[width=\columnwidth]{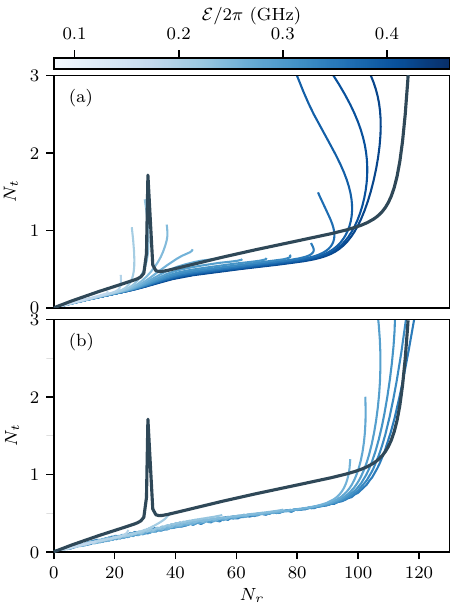}
    \caption{Parametric plots of the average transmon population versus the average resonator population for an  evolution time of $\kappa^{-1}$ for a) $\kappa/2\pi = 20$ MHz as in the main text and b) $\kappa/2\pi = 80$ MHz. Each line represents a unique simulation with a different drive amplitude $\mathcal{E}$ starting in the ground state. The frequency of the drive is set to mean of the pulled resonator frequencies associated to the ground and excited states of the transmon, as is often the case in dispersive readout. For the larger $\kappa$, the system sweeps through the resonance $N_r \sim 30$ at a faster rate. The transition becomes more diabatic, and the leakage through other states is reduced, as expected from the Landau-Zener theory.}
    \label{fig:Parametric_two_kappas}
\end{figure}

\section{Landau-Zener transitions for different parameter choices}
\label{sec:LZ}

\Cref{fig:Parametric_two_kappas}(a) shows the average transmon population versus the average resonator population for the same parameters as in the main text except for a slightly different drive frequency corresponding to the mean of the pulled resonator frequencies associated to the transmon's ground and excited states. This different drive frequency only leads to a small quantitative change with respect to \cref{fig:AllbranchesFadedInteractions}. More importantly, \cref{fig:Parametric_two_kappas}(b) is obtained with that same drive frequency but now a larger decay rates $\kappa/2\pi = 80$ MHz. The resonance around $N_r \sim 30$ mentioned in \cref{section:identification} leads here to smaller leakage. This is compatible with Landau-Zener theory since with a larger $\kappa$ the photon number rises more rapidly corresponding to a faster sweeps through this resonance and therefore a more diabatic transition. 

In addition, we note that the numerically observed slow increase of the transmon population in \cref{fig:Parametric_two_kappas}(b) is overestimated by the semiclassical model. This is not the case for \cref{fig:AllbranchesFadedInteractions}(a) where the drive frequency is set to the bare resonator frequency. We suspect that the following mechanism is responsible for this discrepancy: for the drive frequency used in~\cref{fig:Parametric_two_kappas}, we expect the cavity state to evolve to two very distinct coherent states whether one initializes the transmon in the ground or the excited states. As it was highlighted in \textcite{Leroux-PRApp-2021}, the distance between the resulting polaronic states are likely to diminish mixing of the states. This effect is, however, not accounted for by the semiclassical model since the branches $\{i\}$ are derived from the static spectrum where the drive frequency is not involved.

\end{document}